\documentclass[usenatbib,aas_macros]{mn2e}

\usepackage{graphicx,graphics}
\usepackage{amsmath}
\usepackage{color}

\def\nbody{NBODY6 }
\def\nbodypp{NBODY6++ }
\def\nbodyppgpu{NBODY6++GPU }
\def\nbodygpu{NBODY6-GPU }
\def\nb{$N$-body }

\title[NBODY6++GPU: Ready for the million-body problem]{NBODY6++GPU: Ready for the gravitational million-body problem}  
\author[L. Wang et al.]{Long Wang$^{1,2}$\thanks{E-mail:long.wang@pku.edu.cn}, Rainer Spurzem$^{3,4,5,1}$, Sverre Aarseth$^{6}$, Keigo Nitadori$^{7}$,
\newauthor{Peter Berczik$^{3,4,5,8}$, M.B.N. Kouwenhoven$^{1,2}$, Thorsten Naab$^{9}$}\\
  $^{1}$Kavli Institute for Astronomy and Astrophysics, Peking University, Yiheyuan Lu 5, Haidian Qu, 100871, Beijing, China\\
  $^{2}$Department of Astronomy, School of Physics, Peking University, Yiheyuan Lu 5, Haidian Qu, 100871, Beijing, China\\
  $^{3}$National Astronomical Observatories and Key Laboratory of Computational Astrophysics, Chinese Academy of Sciences, \\
  20A Datun Rd., Chaoyang District, 100012, Beijing, China\\ 
  $^{4}$Key Laboratory of Frontiers in Theoretical Physics, Institute of Theoretical Physics, Chinese Academy of Sciences, \\
Beijing, 100190, China \\
  $^{5}$Astronomisches Rechen-Institut, Zentrum f\"ur Astronomie, University of Heidelberg, M\"onchhofstrasse 12-14, \\
69120, Heidelberg, Germany\\
  $^{6}$Institute of Astronomy, University of Cambridge, Madingley Road, Cambridge CB3 0HA, UK \\
  $^{7}$RIKEN Advanced Institute for Computational Science‏, Kobe, Japan‏\\
  $^{8}$Main Astronomical Observatory, National Academy of Sciences of Ukraine, 27 Akademika Zabolotnoho St., 03680, \\
Kyiv, Ukraine\\
  $^{9}$Max-Planck Institut f\"ur Astrophysik, Karl-Schwarzschild-Str. 1, D-85741 Garching, Germany \\
}

\begin{document}

\date{Accepted 2015 April 9. Received 2015 April 9; in original form 2015 January 9}

\volume{450} \pagerange{4070-4080} \pubyear{2015}

\maketitle

\label{firstpage}

\begin{abstract}
Accurate direct \nb simulations help to obtain detailed information about the dynamical evolution of star clusters.
They also enable comparisons with analytical models and Fokker-Planck or Monte-Carlo methods. 
\nbody is a well-known direct \nb code for star clusters,
 and \nbodypp is the extended version designed for large particle number simulations by supercomputers. 
We present NBODY6++GPU, an optimized version of \nbodypp with hybrid parallelization methods (MPI, GPU, OpenMP, and AVX/SSE) to accelerate large direct \nb simulations, and in particular to solve the million-body problem. 
We discuss the new features of the \nbodyppgpu code, benchmarks, as well as the first results from a simulation of a realistic globular cluster initially containing a million particles.
For million-body simulations, \nbodyppgpu is $400-2000$ times faster than \nbody with 320 CPU cores and 32 NVIDIA K20X GPUs. 
With this computing cluster specification, the simulations of million-body globular clusters including $5\%$ primordial binaries require about an hour per half-mass crossing time.

\end{abstract}


\section{Introduction}

Direct simulations of star clusters have a long history. 
As algorithms and hardware have improved, larger numbers of stars could be simulated, allowing a more realistic representation of the dynamical evolution of globular star clusters. 
\nbody \citep{Aarseth2003} is a state-of-the-art direct $N$-body simulation code specifically designed for star clusters. 
It uses several algorithms to enhance the computing speed and accuracy, especially for strong interactions that arise from a large fraction of binaries and relatively short relaxation timescales ($\le 100$ Myr for typical open clusters and $\le 1$ Gyr for typical globular clusters) in collisional and dense stellar systems.
Here, the terms ``collisional'' and ``dense'' are not well defined in the literature. 
The classical two-body relaxation time, as defined e.g. by \cite{Chandrasekhar1942,Spitzer1987}, describes how important distant gravitational two-body encounters are for the orbital motion of stars. 
If the relaxation time is very long, a system is denoted as ``collisionless'' (for example, galactic disks or bulges); 
the motion of stars is entirely determined by the smooth mean gravitational field of the system. 
If the relaxation time is short (e.g., shorter than the lifetime of the system) we denote the cluster as ``collisional'' (e.g., globular and open star clusters, nuclear star clusters).
If the stellar density is high enough, close two-body gravitational encounters and stellar collisions may occur.
This aspect is crucial when studying ``dense'' star clusters.
In dense and collisional star clusters a correct integration of stellar motions requires pairwise gravitational interactions to be included between many if not all stars in the cluster.
This is the situation for which codes such as \nbody are designed.

Direct \nb simulation of star clusters can be very time consuming. 
In a system with $N$ particles, the full force calculation cost of one particle scales with $O(N)$. 
With individual time steps for each particle, the cost per crossing time ($t_{\rm cr}$) depends on the number of steps per particle ($N_{\rm s}$) which varies with different time step criteria, integration methods and star cluster properties.
\cite{Makino1988} and \cite{Makino1992} found that for the Hermite scheme with a time step criterion based on relative force change \citep{Aarseth1985}, $N_{\rm s}$ is roughly proportional to $N^{1/3}$ for systems with homogeneous density. 
Thus, when using individual time steps the total computational cost per crossing time of $N$ particles scales with $O(N^{7/3})$. 
For systems with a power-law density distribution $\rho \propto r^{-\alpha}$, $N_{\rm s}$ depends on the power index $\alpha$.
Then the cost per crossing time scales with $O(N^{7/3})$ for $\alpha < 24/11$ and $O(N^{(6-\alpha)/(6-2\alpha)})$ for $\alpha \ge 24/11$ \citep{Makino1988}.
Considering the half-mass relaxation timescale, $t_{\rm rh}$ is proportional to $t_{\rm cr} N/\ln{N}$ \citep{Spitzer1987,Sugimoto1990}, the computational cost per $t_{\rm rh}$ is $O(N^{10/3}/\ln{N})$ for homogeneous systems and for power-law systems with $\alpha < 24/11$ and $O(N^{(12-3\alpha)/(6-2\alpha)}/\ln{N})$ for $\alpha \ge 24/11$. 
Thus, an efficient parallelization of a direct \nb code is necessary for large particle numbers.

\cite{Sugimoto1990} discussed the fundamental problem that direct numerical simulations of globular star clusters could not be completed for decades if extrapolating the standard evolution of computational hardware (Moore's law).
They called for the construction of a special-purpose computer GRAPE, which finally was successfully initiated and completed by their team \citep{Makino1993,Makino1998,Makino2003}.
In the following years, graphical processing units (GPU) widely replaced GRAPE \citep[e.g.,][]{Harfst2007,Hennebelle2007,PZ2007,Belleman2008,Schive2008} and much of the GRAPE software could be ported to GPU \citep{Gaburov2009}.

\cite{Spurzem1999,Spurzem2008} and \cite{Hemsendorf2003} discussed several different types of hardware for parallelization and extended \nbody to \nbodypp  for general parallel supercomputers.  
Later, \cite{Nitadori2012} developed a GPU-based parallel force calculation for NBODY6.
As a result, large $N$-body simulations ($N\sim 10^5$) became possible on a single desktop computer or workstation with GPU hardware. 
They also implemented the parallel force calculation based on Streaming SIMD Extensions (SSE) and Advanced Vector Extensions (AVX) for recent CPU architectures.
\cite{Spurzem2011,Berczik2013a} and \cite{Berczik2013b} discussed the performance of large \nb simulations with the GPU-accelerated codes $\phi$GPU and a provisional version of NBODY6++GPU.

With these parallelization methods, we can now study star clusters with a number of stars exceeding $10^5$. 
\cite{Hurley2012} simulated $200,000$ stars including $5000$ primordial binaries with initial half-mass radius $4.7$~pc using NBODY4 on a GRAPE-6 based computer to investigate core collapse and core oscillation.  
Later, \cite{Sippel2013} studied the multiple stellar-mass black holes in globular clusters by simulating $262,500$ stars including $12,500$ primordial binaries with initial half-mass radius $6.2$~pc using NBODY6-GPU.
The current largest direct \nb simulation modeling the globular cluster M4 used one computing node including 12 Intel Xeon X5650 cores (2.66 GHz per core) and 2 NVIDIA TESLA C2050 GPUs with 448 cores each (1.15 GHz per core) with \nbodygpu \citep{Heggie2014}. 
This simulation contained $~500,000$ stars with $7\%$ binaries and a small half mass radius of $0.58$ pc. 
Nowadays, we can make an effort to reach one million stars by using parallel supercomputers with GPUs.

In this paper, we first introduce the parallel algorithm used by \nbodygpu and \nbodypp in Section~\ref{sec:feature}. 
Then we describe the new version of \nbodyppgpu with a hybrid parallel method and also the new algorithms that are necessary for large number of particle parallelization in Section~\ref{sec:new}. 
Performance tests are carried out in Section~\ref{sec:performance}.
In Section~\ref{sec:app} we show an application to a globular cluster with one million stars. 
In Section~\ref{sec:discussion}, we discuss the parallelization limit and future development of NBODY6++GPU. 
Finally, we present our conclusions in Section~\ref{sec:conclusion}.

\section{The features of NBODY6/6++}
\label{sec:feature}
\nbody uses the fourth-order Hermite integration method.
\cite{Makino1991} presented a careful analysis of the performance and energy error of the Hermite integrator. He showed that it reduces to the similar asymptotic error behaviour as the standard Aarseth scheme (fourth-order method; see \citealp{Aarseth1985}) but it has some advantages in the time step choice and data structure.

The hierarchical block time steps method is used together with the Hermite integrator \citep{McMillan1986,Makino1991} in NBODY6, 
which avoids the overheads of particle position and velocity prediction in an individual time step method. 
In this method, particle time steps are adjusted to quantized values, usually an integer power of $0.5$. 
Then at each time step, active particles (the particles that satisfy the time step criterion) are integrated together.

To speed up the force calculation, \nbody uses the Ahmad-Cohen (AC) neighbor scheme \citep{Ahmad1973}.
The basic idea is to employ a neighbor list for each particle. 
The integration is separated into two parts: regular force integration for large time steps (regular steps) and irregular force integration for small time steps (irregular steps). 
The regular force is the summation of the forces from particles outside the neighbor radius and the irregular force accumulates only the neighbor forces. 
During the irregular step, the regular force and its first order derivative calculated at the last regular step are used for position and velocity prediction.
The AC scheme gains efficiency with sequential computing (without parallelization).
The speed gained by the AC scheme is roughly proportional to $N^{1/4}$ \citep{Makino1988,Makino1992}.
However, in parallel computing, this gain is limited by the complexity of the implementation of this algorithm (see Section~\ref{sec:performance}). 
Also, the benefit of reducing overheads of particle prediction in the block time step method is strongly limited in the neighbor scheme.

\cite{PZ2014} discussed the integration accuracy requirement for self-gravitating systems simulated with direct \nb codes.
They found that for three-body systems the integration should have total energy conserved better than $1/10{\rm th}$. 
Although this accuracy requirement is uncertain when the simulation is extended to large particle number systems, this work indicates the importance of careful integration treatment for direct \nb systems.
One important feature of \nbody is that it uses the algorithms of \cite{Kustaanheimo1965} (hereafter KS) and chain regularization \citep{Mikkola1993} to deal with an accurate solution of close encounters, binaries and multiple systems, which play a significant role in star cluster dynamical evolution.
These strong interactions require very small time steps during integration and may produce large errors with standard integrators such as the Hermite scheme. 
Using KS and chain regularization is also the most important feature of \nbody for star cluster simulations.

Here, we have briefly introduced the main algorithm used in NBODY6/6++.
In the next section we will focus on the parallelization of the codes.

\section{Parallelization of NBODY6++GPU}
\label{sec:new}

\subsection{MPI parallelization of NBODY6++}

\cite{Spurzem1999} and \cite{Hemsendorf2003} developed \nbodypp based on \nbody using MPI parallelization with the copy algorithm.  
Both regular and irregular forces were parallelized. 
Here different MPI processors calculate different subsets of the active particles. 
Each MPI processor has the complete particle dataset. 
Another available parallel algorithm is the ring algorithm which splits the full particle dataset for different MPI processors. It reduces the memory cost in each MPI process. 
The benefit of the copy algorithm compared to the ring algorithm is that there is no requirement for extra communication of the neighbor particle data which is not in the same MPI process during the irregular force calculation. 
The disadvantage is the particle number limit due to memory size on the computing node.  
The MPI communication with the copy algorithm has constant time cost (independent of MPI processor number except for latency).
The scaling of the regular force with different MPI processors is very good. 
Since the regular force dominates the calculation, this results in a good scaling of the total computing time.
\cite{Dorband2003} provided a detailed discussion of these communication algorithms. 
\cite{Lippert1998} and \cite{Makino2002} suggested an efficient communication algorithm (hypersystolic) for extremely large processor numbers.

\subsection{Basic NBODY6-GPU implementation}
After the GPU computing (CUDA) became popular, the shared memory parallel \nbodygpu code was developed for workstation and desktop computers \citep{Nitadori2012}. 
The OpenMP, GPU (CUDA) and AVX/SSE parallel methods are used to make the code as fast as possible.
However, \nbodygpu can only be used in a single node (no massively parallel MPI implementation) so the number of particles is limited for a reasonable simulation time.

\subsubsection{Regular force and potential (GPU)}
The GPU library of \cite{Nitadori2012} is used for calculating the regular force, which dominates the direct integration, and potential energy calculation. 
The cost for regular force calculation per particle scales with $O(N)$ and for potential energy calculation scales with $O(N^2)$. 
The performance of GPU force calculation is very good since the pure force calculation is easy to parallelize. 
GPUs also help to accumulate the neighbor list very efficiently during the regular force calculation.

\subsubsection{Prediction and irregular force (AVX/SSE)}
When GPU accelerates the regular force very efficiently, the irregular force becomes expensive. 
However, this part is hard to parallelize on GPUs due to the complexity of the AC neighbor scheme. 
Thus, \cite{Nitadori2012} developed the AVX/SSE and OpenMP parallel library for neighbor particle prediction and irregular force calculation.
AVX/SSE is an instruction set for CPUs developed in recent years, which supports vector calculation in the specific cache. 
The advantage of AVX/SSE with OpenMP is that there is no extra memory copy compared to GPU.
For both AVX/SSE and GPU libraries, the data needs to be copied once for changing data structure to obtain computing efficiency.
This is because that \nbody has a very long development history, thus to completely change the data structure to be consistent with AVX/SSE and GPU libraries is very time consuming.
But for GPU, there is extra data copy from the host memory on the mother board to the device memory on GPU.
Besides, since the neighbor force calculation is not efficient for the distributed memory parallel method (with MPI parallelization; see discussion below), 
this kind of shared memory parallel method is more efficient.

\subsection{Code improvements in NBODY6++GPU}

In this subsection we describe our new implementations in NBODY6++GPU.

The GPU acceleration, especially of the long-range (regular) gravitational forces, is very efficient so this part does not dominate the computational time any more, as we show below. 
Secondly, the AVX/SSE implementation accelerates prediction and neighbor (irregular) forces, which is the next most time consuming part of the code. 

We have combined the GPU and AVX/SSE acceleration, which was done for a single node in NBODY6-GPU, with the MPI parallelized NBODY6++ designed for multi-node computing clusters for the new version NBODY6++GPU.
This work requires additional efforts to keep the code consistent (see below). 
In addition, we have worked on remaining bottlenecks, such as time step scheduling and stellar evolution, which become important for million bodies because the usual computationally intensive tasks have been accelerated very effectively by GPU and AVX/SSE.

\subsubsection{New algorithm of selecting active particles for block time steps}
\label{sec:initb}

\begin{figure}
  \centering
  \includegraphics[width=0.5\textwidth,height=!]{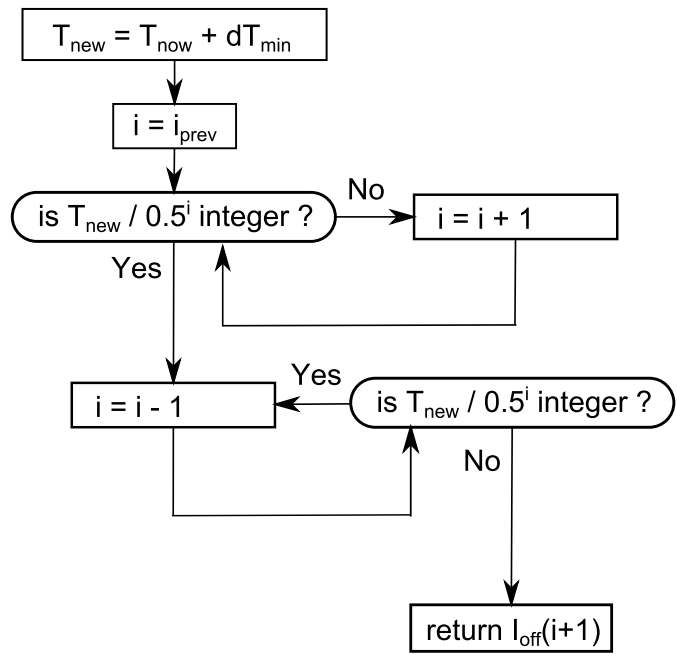}
  \caption{The flow chart for obtaining the correct offset in the sorting list algorithm. 
    $T_{\rm new}$ is the time after next integration. $T_{\rm now}$ is current time. $dT_{\rm min}$ is the current smallest time step (the time step of the first particle in sorting list).
    $i_{\rm prev}$ is the previous offset. }
  \label{fig:sortchart}
\end{figure}

\begin{figure}
  \centering
  \includegraphics[width=0.5\textwidth,height=!]{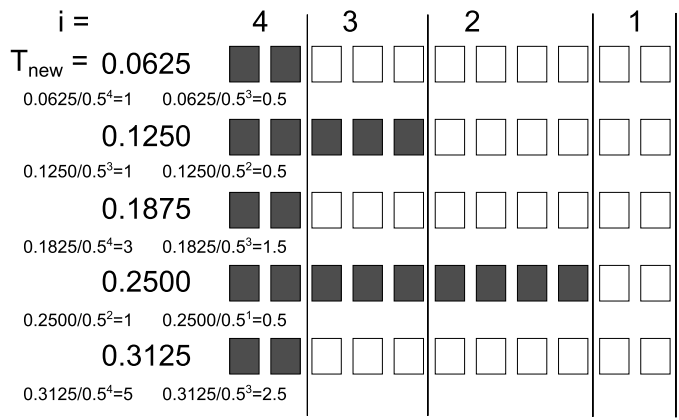}
  \caption{Diagrammatic sketch of sorting list algorithm for selecting active particles. 
    The time step of each particle block is $0.5^i$ separated by boundary indicators $I_{\rm off}(i)$ (vertical lines).
    The integration advances vertically in the chart. Active particles are shown as black squares.}
  \label{fig:sortlist}
\end{figure}

For the block time step method, active particles should be selected at every time step. 
It is very expensive to search all particles for the active ones, especially for the irregular force calculation. 
In this case, for one block time step the cost of selecting active particles scales with $O(N)$ while the irregular force calculation cost scale with $O(N_{\rm i} \langle N_{\rm b} \rangle)$. 
If $N \gg N_{\rm i} \langle N_{\rm b} \rangle$, the former can be more expensive.
When the simulation reaches millions of particles, the block time step levels can be quite deep (the smallest time step can reach $0.5^{20}-0.5^{22}$) and the deep blocks with few particles and small time steps can easily satisfy this condition. 
One may consider to use a temporary list to save particles with small time steps and only search all particles at some selected time interval from the temporary list each time step.
However, this method is still expensive where there are many particles with small time steps (such as the wide binaries that are not KS regularized). 
Indeed, we find that the time of selecting active particles can be much larger than the irregular integration time, even with this temporary list algorithm for one million particles including $5\%$ primordial binaries. 
Another reason that forces us to deal with this issue is that the active particles selection is very difficult to parallelize efficiently (the cost is almost independent of processor numbers) and would be prohibitive for a million-body simulation.
Thus, we propose a better algorithm that uses a time step sorting list (hereafter sorting list algorithm; see Figures~\ref{fig:sortchart} and \ref{fig:sortlist}). 
\cite{Zhong2014} implemented a similar algorithm for $\phi$-GRAPE+GPU and evaluated its performance.
The basic idea is that when we have the index list sorted by particle time step from smallest to largest, and the indicators of each boundary offset $I_{\rm off}(i)$ between the block of the same step particles (the largest particle index with step $0.5^i$), 
we only need to find the correct offset at each block time step by using the algorithm shown in Figure~\ref{fig:sortchart} to select active particles (shown as black squares in Figure~\ref{fig:sortlist}). 
After integration, we adjust the sorted list by sorting the active particles' new time steps.
The specific sorting method for this adjustment can be optimized to $O(N_{\rm i})$ 
if we ignore the stability of sorting (stability means no exchange of the order for the particles with same steps) and assume that many active particles keep the same step as before or have small time step changes.

\subsubsection{ The initialization }
\label{sec:init}

The initialization of a simulation in \nbody is relatively expensive. 
We improve it with MPI, GPU and OpenMP parallelization and a better algorithm.
The initial model for million-body simulations is very important and needs to be carefully tested.
This improvement is very useful for fast testing of the initial models with large particle numbers, especially for a large number of primordial binaries.

The initialization of \nbody can be divided into four parts:
\begin{enumerate}
\item reading or generating masses, positions, velocities and stellar evolution parameters of all stars;
\item scaling all parameters into \nb units (the \nb units\footnote{It has been suggested to name the \nb time unit to honour M. H$\acute{e}$non as H$\acute{e}$non time unit (D.C. Heggie, private communication)} are defined in \citealp{Heggie1986});
\item initialization of forces, neighbor lists and time steps of all stars;
\item initialization of primordial KS binaries.
\end{enumerate}
In the second part of the intialization, the total potential energy of the system is needed and costs $O(N^2)$. 
Actually, \nbodygpu does this calculation twice for scaling purpose in the case of an external tidal field.
The GPU is used in \nbodygpu to speed up this part and it is very efficient. 
Our new improvements are for the third and fourth parts. 
In the traditional \nbodygpu version forces and neighbor lists are initialized separately without parallelization. 
NBODY6++ parallelizes the scaling and initialization of the force parts, but only through MPI.
For million-body simulations this is very slow and requires hours to be finished.
We improved it by using GPU based force and neighbor list calculations (the same as for the regular force calculation). 
The fourth part is very costly with more than $5\%$ primordial KS binaries in the traditional \nbodygpu (several hours). 
During initialization of KS binaries, the force and its three derivatives (Hermite scheme) need to be renewed for center-of-mass particles. 
All neighbor lists that contain KS binary component indices also need to be replaced by the center-of-mass particle indices. 
The cost is approximately $O(N \langle N_{\rm b} \rangle N_{\rm KS})$ where $N_{\rm KS}$ is the number of primordial KS binaries. 
We find a much simpler way to initialize KS binaries (cost scales with $O(N_{\rm KS})$) by just switching the order of the third and fourth parts: 
initialize KS binaries first without recalculating forces, their derivatives and neighbor lists (only the KS transformation is needed) and then do the third process with the new center-of-mass particles data generated by former process instead of each binary component in the old way. 
In this case there is no need to update the forces and neighbor lists. 

\subsubsection{Position and velocity prediction}

During the force calculation, the predicted positions and velocities are used to calculate the force and its first derivative for the Hermite integrator. 
In principle, we can avoid prediction of the same particles with the AC neighbor scheme and block time steps. 
However, in practice we need to search all neighbors of each active particle and the search itself is computationally expensive. 
Thus, it does not save much time to avoid neighbor prediction overlap and it is much simpler to predict all neighbors and do the force calculation within one loop. 
The disadvantage of this method is that it costs more when the average neighbor number $\langle N_{\rm b} \rangle$ multiplied by the active particle number $N_{\rm i}$ is larger than the total particle number $N$,
compared to all the particle predictions with a non-AC scheme block time step. 
One solution is to try predicting all particles once instead of predicting each neighbor when $\langle N_{\rm b} \rangle N_{\rm i} > N$. 
But the mixture of predicting only neighbors and predicting all particles increases the complexity of code. 
We therefore use only neighbor prediction in the code.

However, there is a major complication for the parallel neighbor prediction in NBODY6++GPU, which does not exist in NBODY6-GPU.
Since we use AVX/SSE and GPU and the code is mixed with \texttt{CUDA}, \texttt{C++} and \texttt{Fortran~77} programming language, the AVX/SSE and GPU libraries keep the individual copies of particle datasets. 
Thus, the predictions of particles have overlaps and are usually inconsistent for different copies distributed on MPI processors.
Due to the complexity of NBODY6/6++ (e.g., using predicted positions for regularization) this leads to problems of synchronization later on, such as differences of time steps for the same particle on different processors.
The safest but very costly way is to always predict all particles at every irregular integration step, which is the case in the older versions of NBODY6++.
To solve this problem, much effort has been made to ensure that every particle is predicted to the current time before it is used in stellar evolution, KS and hierarchical regularization, because these parts are not parallelized and should have the same computing results on every MPI processor.

\subsubsection{ Stellar evolution and neighbor force correction}
The neighbor scheme also leads to performance losses for the calculation of stellar evolution. When a star experiences mass loss, other stars feel a smaller force. 
In the neighbor scheme, the regular force is predicted from the value calculated at the last regular time step, 
thus if particles outside the neighbor radius experience mass loss between the previous and next regular time steps, the regular force will be inconsistent after that. 
The correction for the regular force should be done for all particles which have the mass loss particle outside their neighbor radius. 
To avoid a large value of the third and fourth derivatives of the force, the irregular force also needs to be updated if the mass loss particle is inside the neighbor radius. 
When mass loss is frequent, the calculation performance will be reduced significantly. 
We currently use OpenMP to speed up the force correction, but it cannot completely solve this issue since the force correction with cost of $O(N)$ per particle cannot be avoided. 

\subsection{ Hybrid MPI parallelization}

\begin{figure*}
  \centering
  \includegraphics[width=1.0\textwidth,height=!]{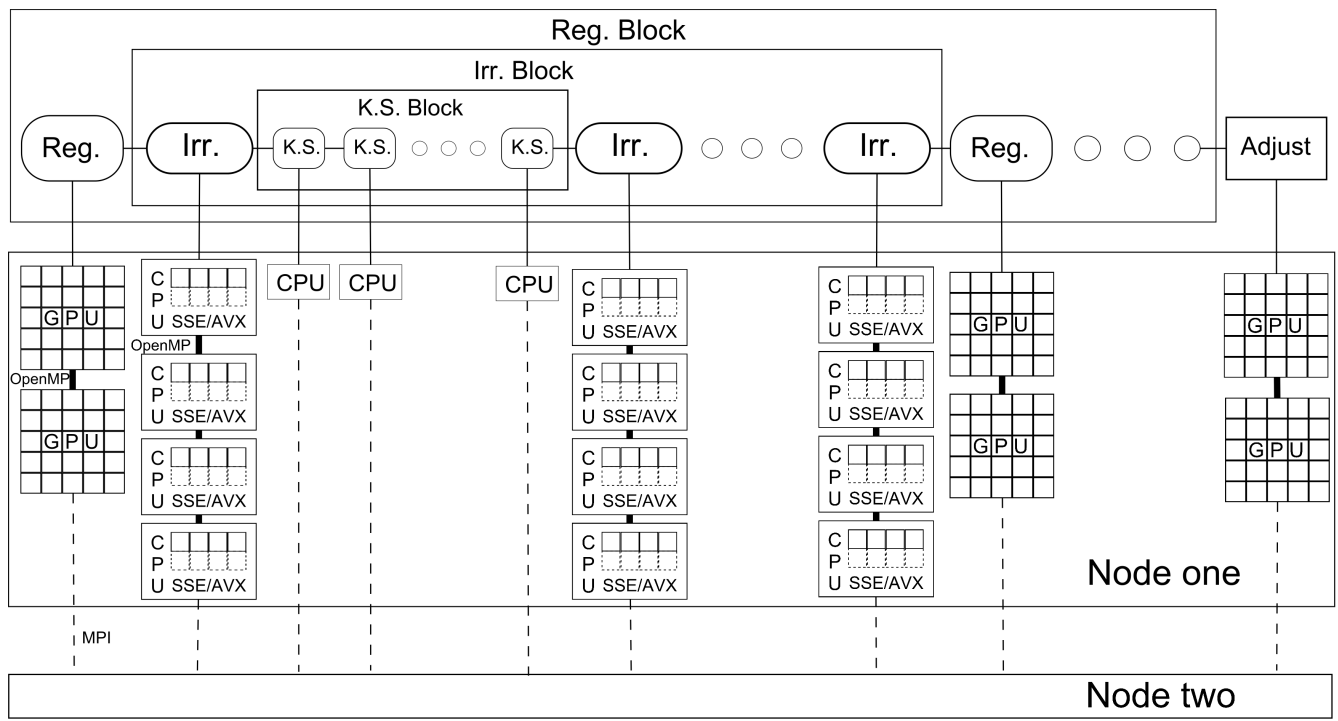}
  \caption{\nbodyppgpu code structure. It shows one cycle of simulation. Based on the time steps, the integration can be divided into three hierarchical parts (see Table~\ref{tab:def}): KS calculation (KS), irregular integration (Irr.) and regular integration (Reg.). 
The KS has smallest time step distribution. 
Thus, between two nearest Irr. block time steps there are several KS steps.
Similarly, between two Reg. block time steps there are several Irr. time steps.
After several Reg. time steps there is one ``Adjust'' (see Table~\ref{tab:def}). 
Inside one node, Reg. and Adjust are parallelized by multiple GPUs and Irr. is parallelized by AVX/SSE with OpenMP. 
MPI parallelization are done for all 4 parts between different nodes.
}
  \label{fig:structure}
\end{figure*}

Based on the above parallel methods, we develop a new version of \nbodyppgpu to include hybrid parallel procedures. 
The parallel structure of \nbodyppgpu is shown in Figure~\ref{fig:structure}. 
In computer clusters, each computing node uses one MPI process. 
Each MPI process opens multiple threads via OpenMP for the irregular force calculation. 
GPUs inside one node are controlled by OpenMP threads. 
Each GPU has a similar particle dataset size for regular force and potential energy calculation. 
GPUs of different nodes are isolated without communication. 
Thus all GPUs in the same node together access the complete particle dataset. 
The best code configuration is to use multiple CPU cores (such as $8-16$ cores) and several GPUs (such as $1-4$ GPUs with a few thousand cores) per node, and choose node numbers based on the total number of particles.

\begin{table*}
  \caption{The definitions of abbreviations for all figures}
  \begin{tabular}{l|l}
    \hline
    Abbreviation & Definition \\\hline
    Reg. & Regular integration (force) and neighbor list determination \\
    Irr. & Irregular integration (force, prediction (AVX/SSE version) and correction)\\
    Pred. & Neighbor (Non-AVX/SSE version) and all particles (for regular force) prediction \\
    Init.B & Initialization of active particle list for block time step \\
    Move  & Particle data copy prepared for MPI communication \\
    Comm.I. & MPI communication for irregular integration \\
    Comm.R. & MPI communication for receiving integration \\
    Send.I. & Particle data copy for AVX/SSE irregular force calculation\\
    Send.R. & Particle data copy for GPU regular force calculation \\
    Adjust  & Energy checking, adjustment of parameters and data results \\
    KS      & KS regularization calculation (binary and hierarchical systems) \\
    Barr.   & MPI communication barrier waiting time due to the imbalance and network traffic between different nodes \\
    \hline
  \end{tabular}
  \label{tab:def}
\end{table*}

\section{Performance test}
\label{sec:performance}

\subsection{ Pure MPI and hybrid MPI}

\begin{figure*}
  \centering
  \includegraphics[width=1.0\textwidth,height=!]{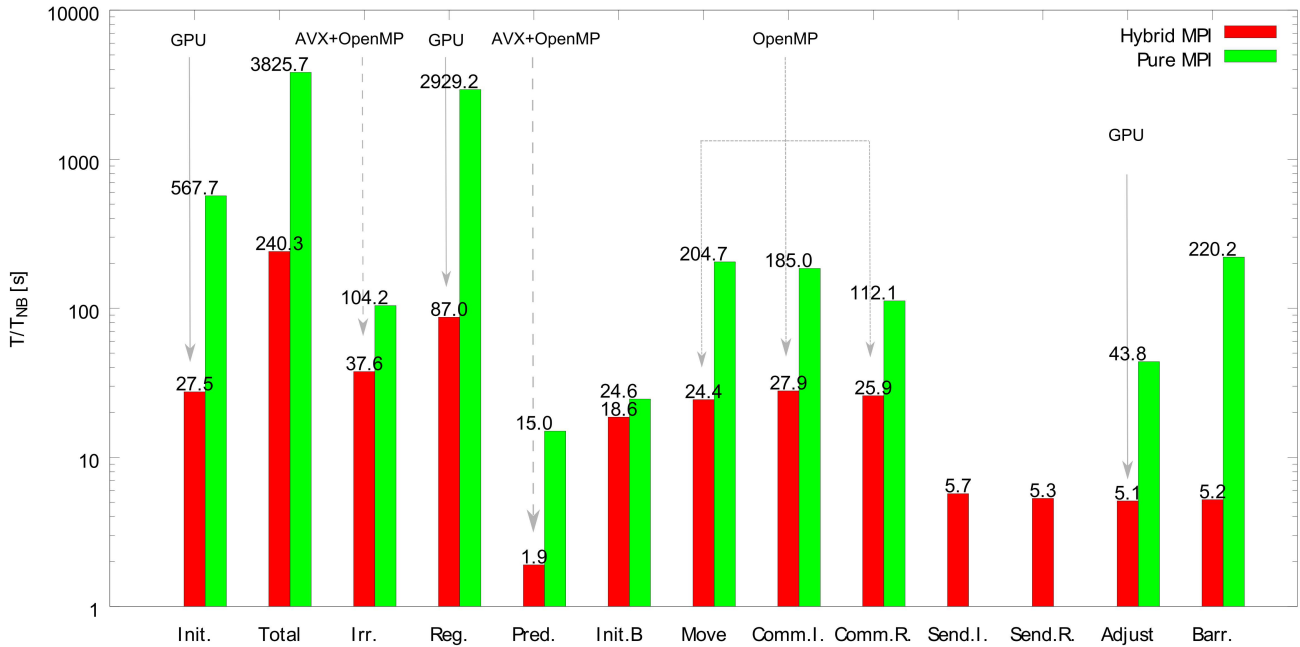}
  \caption{Comparison of performance between pure MPI and hybrid MPI (GPU + AVX/SSE + OpenMP + MPI) on the ``Kepler'' cluster at ARI, Heidelberg University. 
    The test uses 256k particles with a Plummer model, IMF from Kroupa (2001) with mass range $0.08 - 100 M_\odot$. 
    The hybrid MPI test uses $4$ nodes and each node includes $32$ Intel Xeon E5-2650 cores ($2.00$ GHz per core) and $4$ NVIDIA K20m with $2496$ cores each ($706$ MHz per core). 
    The pure MPI test uses the same configuration of nodes and CPU cores. 
    The label ``Total'' means total time cost for $1$ \nb unit and ``Init.'' denotes the initialization time of the simulations. }
  \label{fig:mpivshybrid}
\end{figure*}

Figure~\ref{fig:mpivshybrid} shows significant improvement of \nbodyppgpu by using hybrid MPI including GPU, as compared to the pure MPI case. 
We see that the GPU gives about $33$ times faster regular force integration.
This is to be expected since GPU is designed for large parallelization by using many computing cores and large memory bandwidth within one card. 
Using AVX/SSE with OpenMP gives about $3$ times faster irregular integration including predictions. 
OpenMP reduces the MPI communication cost by a factor of $5-10$.
The individual MPI communication process is not directly sped up by OpenMP. 
When we use MPI parallelization together with OpenMP method, inside one node the irregular and regular force calculations are done by multiple threads with OpenMP instead of MPI parallelization, thus we can set a larger block particle number threshold for MPI parallelization by a factor of the OpenMP thread number and reduce the total MPI communication frequency. 
This then results in shorter total MPI communication time.

\subsection{Scaling with different particle numbers and processors}
\label{sec:scale}
\begin{figure}
  \centering
  without primordial binaries\\
  \includegraphics[width=0.5\textwidth,height=!]{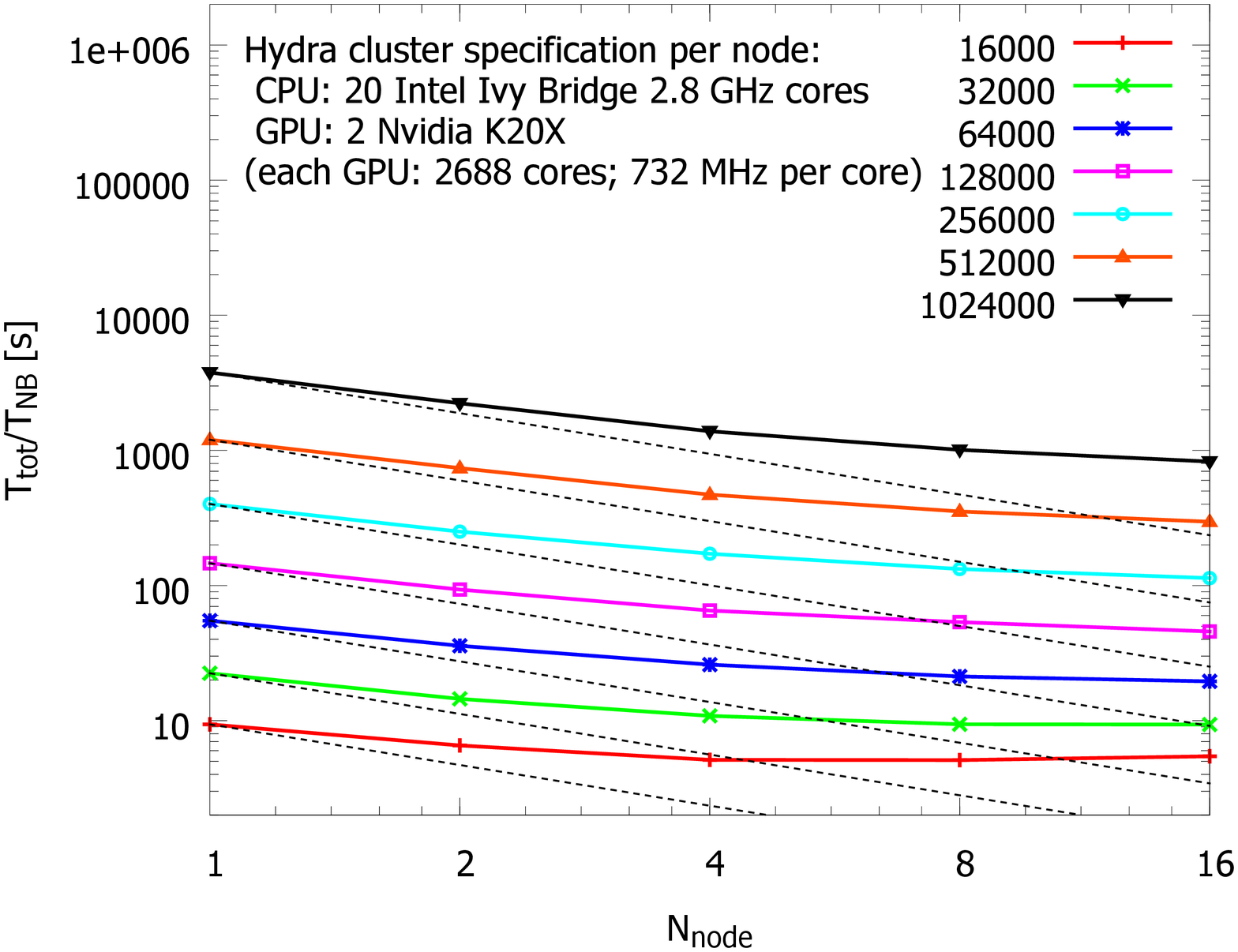}\\
  with $5\%$ primordial binaries\\
  \includegraphics[width=0.5\textwidth,height=!]{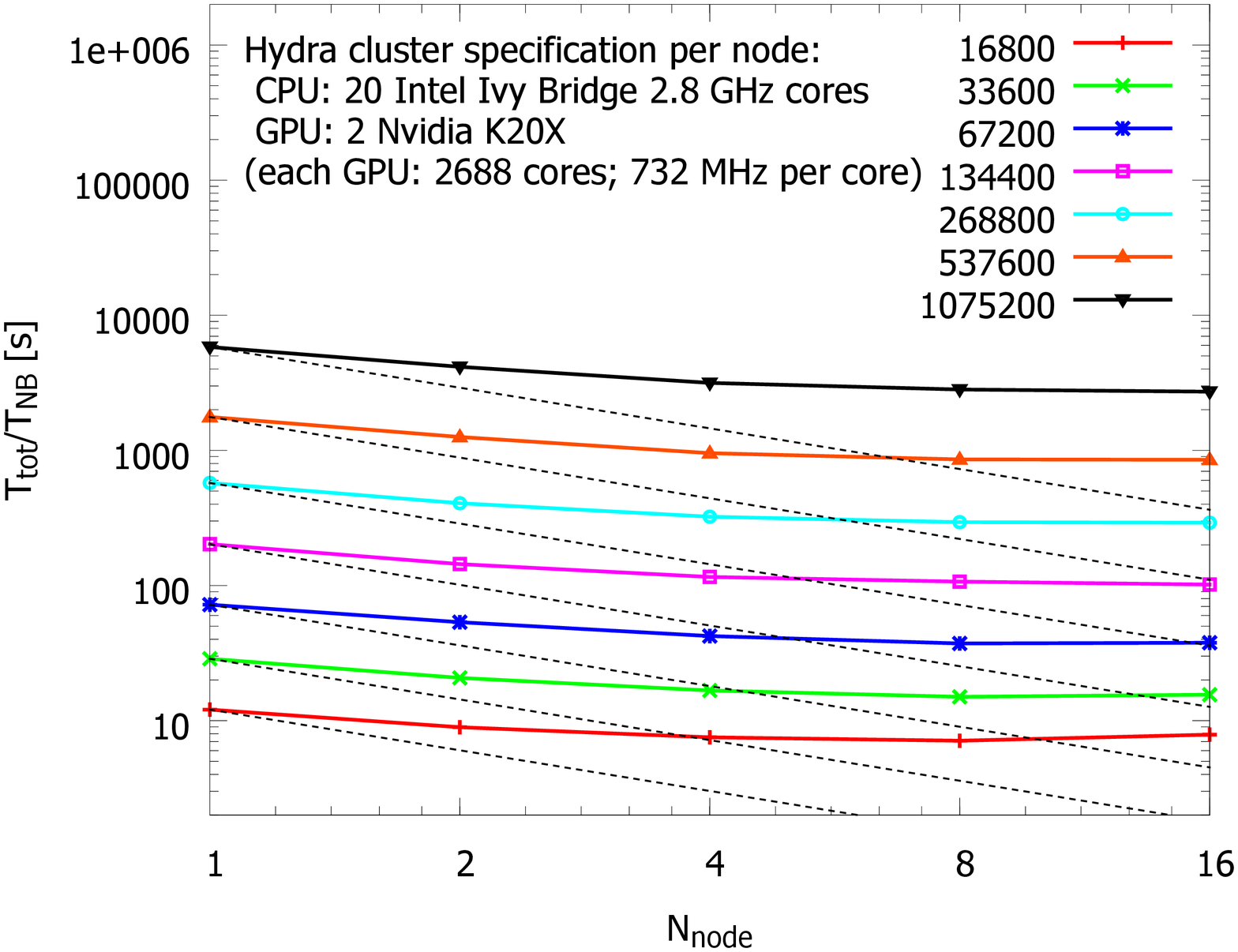}
  \caption{Performance of \nbodyppgpu with hybrid MPI on the ``Hydra'' cluster as the function of node number $N_{\rm node}$. 
    $T_{\rm tot}/T_{\rm NB}$ shows the computing time cost per \nb time unit.
    The configurations of each node are indicated in the panels. The dashed line shows the
    ideal parallel limit with zero communication cost. Different colors represent different particle numbers.}
  \label{fig:scaling}
\end{figure}

\begin{figure}
  \centering
  \includegraphics[width=0.5\textwidth,height=!]{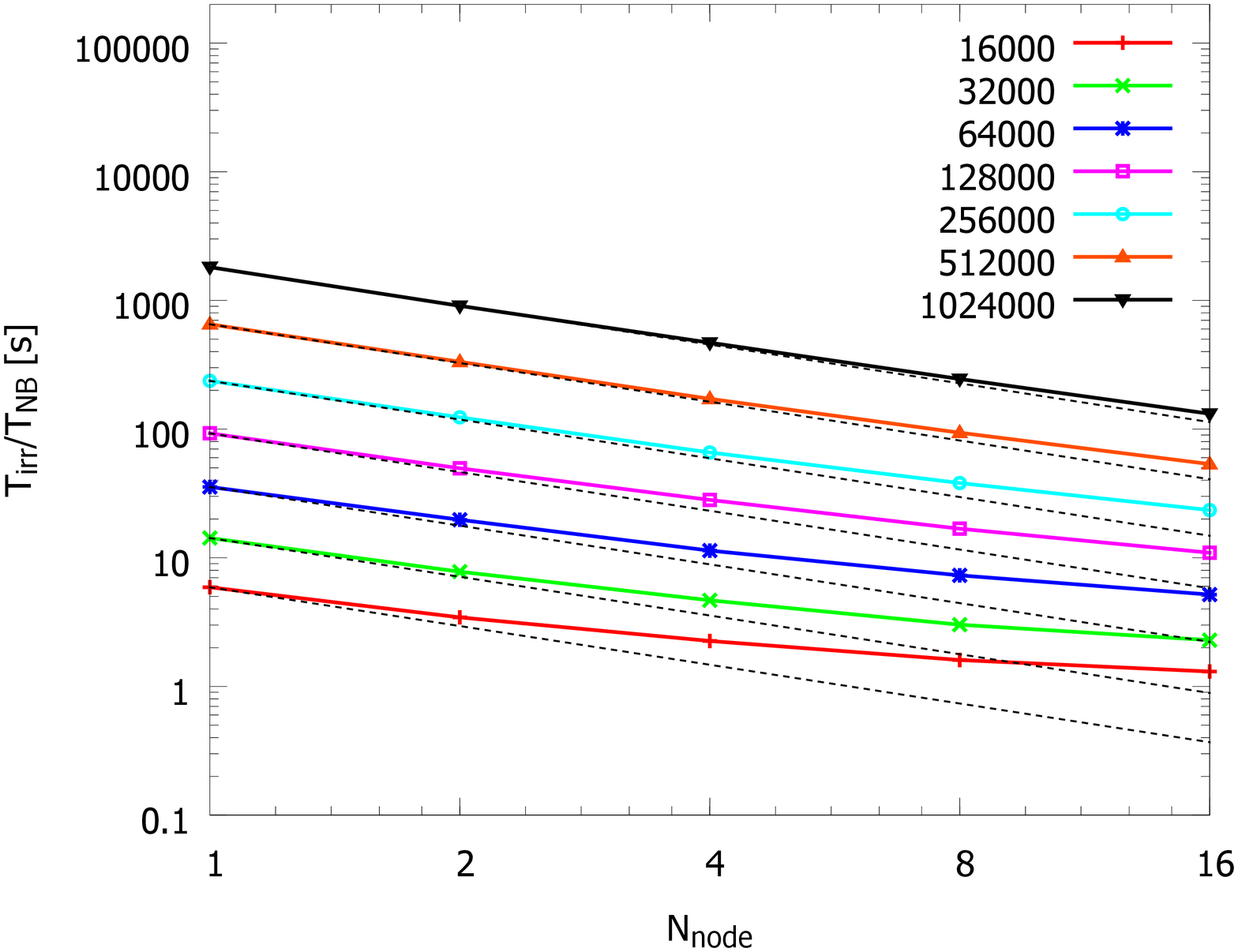}\\
  \includegraphics[width=0.5\textwidth,height=!]{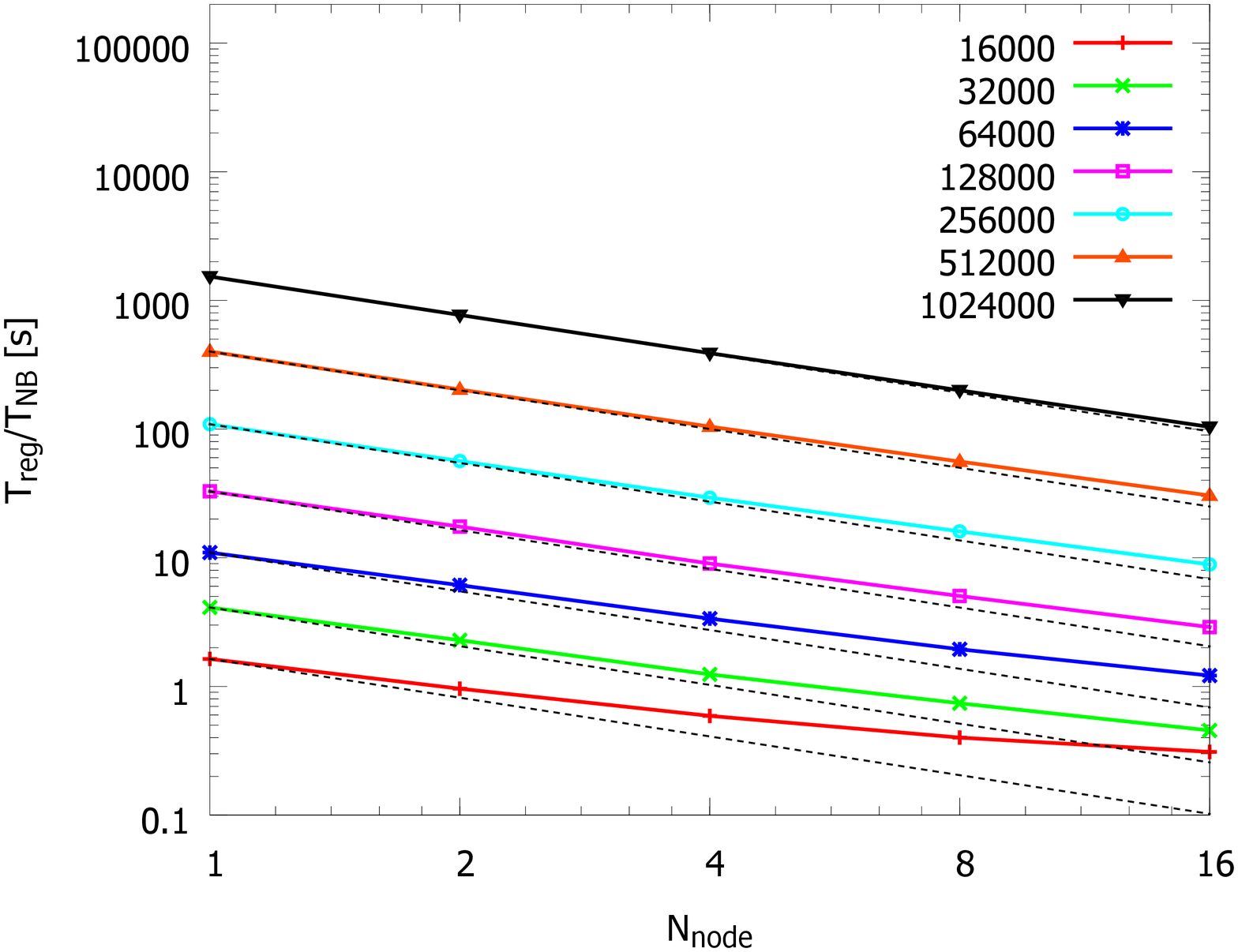}
  \caption{Performance of regular and irregular integration on the ``Hydra'' cluster as the function of $N_{\rm node}$. Here the same node configurations and line types as in Figure~\ref{fig:scaling} are used. $T_{\rm irr}/T_{\rm NB}$ and $T_{\rm reg}/T_{\rm NB}$ shows the irregular and regular integration computing time cost per \nb time unit respectively.}
  \label{fig:regirr}
\end{figure}

The scaling with different particle numbers and processors demonstrate the possibility of using large computing resources for simulations. 
We test hybrid parallel \nbodyppgpu scaling with different node numbers $N_{\rm node}$ (1, 2, 4, 8 and 16; up to 320 CPU cores and about 80k GPU cores) and different particle numbers (16k, 32k, 64k, 128k, 256k and 1024k) on the ``Hydra'' cluster of the Max-Planck Supercomputing Centre (RZG) Germany.
Each node is completely controlled (no other tasks on the node) and has two NVIDIA K20X with 2688 cores each (732 MHz per core) and 20 Intel Ivy Bridge cores (2.8 GHz per core). 
The total computing time for one \nb time unit $T_{\rm tot}/T_{\rm NB}$ is shown in Figure~\ref{fig:scaling}.
The irregular and regular force integration computing time ($T_{\rm irr}$ and $T_{\rm reg}$) are shown in Figure~\ref{fig:regirr}. 
All these three times are the averaged computing times of the first two \nb time units of each simulation.
We test two basic initial models. 
One has no primordial binaries and another has $5\%$ binaries. 
Both use a Plummer sphere \citep{Plummer1911} and initial mass function (IMF) from \cite{Kroupa1993} with mass range $0.08-20 M_\odot$ and no stellar evolution. 

In the non-binary case, the scaling with different $N_{\rm node}$ for the total time is not ideal because of the communication cost. 
Here the speed-up saturates at about $8-16$ nodes depending on the particle number.
But if we consider the number of cores per node (20 CPU cores and 5376 GPU cores), the scaling with cores is excellent, since with $16$ nodes $320$ CPU cores and $86016$ GPU cores are used.
With one node, the performance of NBODY6++GPU is similar to that of NBODY6-GPU. \cite{Nitadori2012} showed that NBODY6-GPU gives about $100$ times speed-up compared to the sequential NBODY6 with two NVIDIA GeForce GTX 560 Ti with 384 cores each (822 MHz per core) and 4 Intel i7-2600K cores (3.40 GHz per core). 
On the ``Hydra'' cluster node, the speed-up can reach $100-500$ depending on the particle number. 
Thus, with $16$ nodes for one million particles, we can reach a factor of $400-2000$ speed-up compared to the sequential NBODY6.
Besides, the absolute time cost is very good, especially for the million-body case, the total time is about $800$~s for $N_{\rm node} = 16$. 
For the $\phi$GPU code tested in the ``Laohu'' cluster with 32 NVIDIA K20 GPU, one million particles take about $1500$~s. 
Although we cannot compare the two codes directly with different computing cluster specifications, with the similar GPU type and number \nbodyppgpu can reach better performance.
CPU cores are ignored here because the time fractions on the CPU for these two codes are very different: $\phi$GPU spends about $90\%$ computing time on the GPU while \nbodyppgpu has much less time fraction (Section~\ref{sec:tfrac}).
In the case with $5\%$ primordial binaries, the scaling is not as good as for the case with no binaries due to the KS calculation. We discuss this issue in more detail in Section~\ref{sec:discussion}.

The regular and irregular integration times in Figure~\ref{fig:regirr} are close to ideal for million-body simulations. 
It means that when ignoring MPI communications, the MPI parallelization speeds up regular and irregular calculation excellently for large number of particles.
For small particle numbers ($< 10^5$) the scalings of both regular and irregular integrations depart from the ideal parallel limit.
The reason is that the number of operations on each node (for regular integration is on GPU) is small. 
Thus, the cost of internal memory accessing and modification during integration, which cannot be scaled with computing cores or nodes, dominates the time.

\subsection{Time fraction for different parts}
\label{sec:tfrac}
\begin{figure}
  \begin{tabular}{cc}
  \includegraphics[width=0.22\textwidth,height=!]{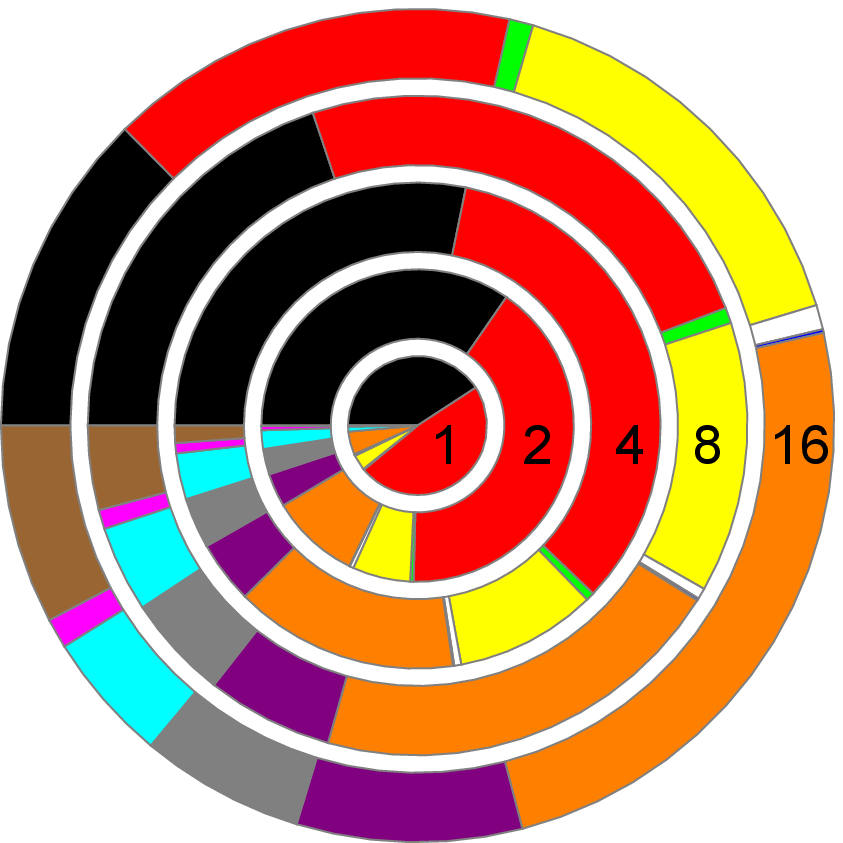} &
  \includegraphics[width=0.22\textwidth,height=!]{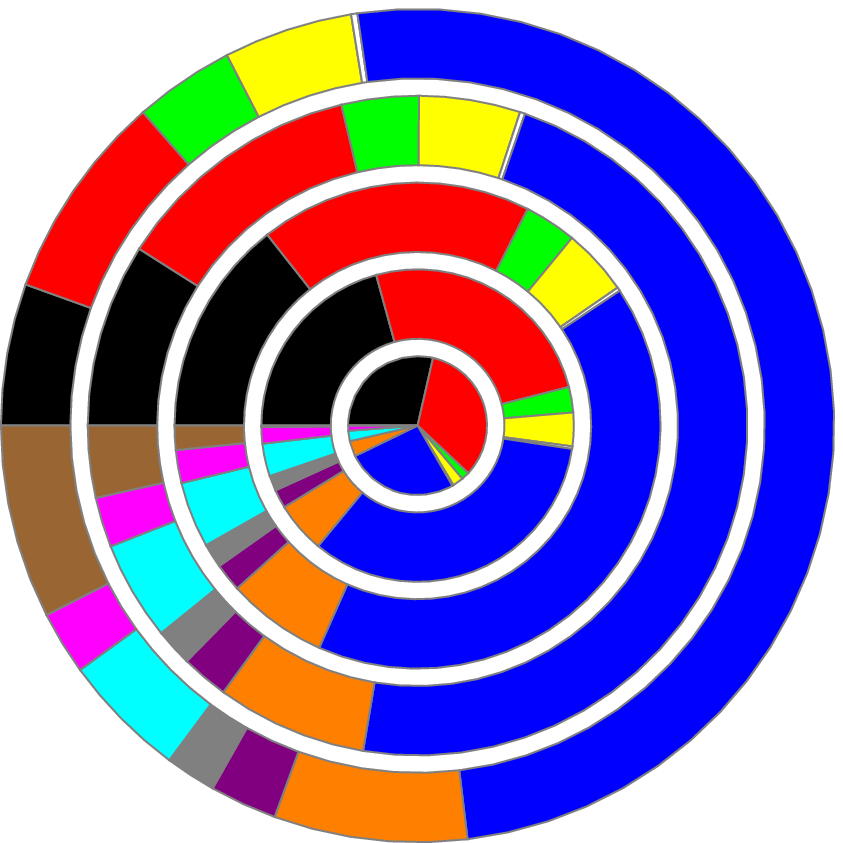} \\
  {\tiny 1024k without primordial binaries} & {\tiny 1024k singles + 51.2k primordial binaries} \\
  \includegraphics[width=0.22\textwidth,height=!]{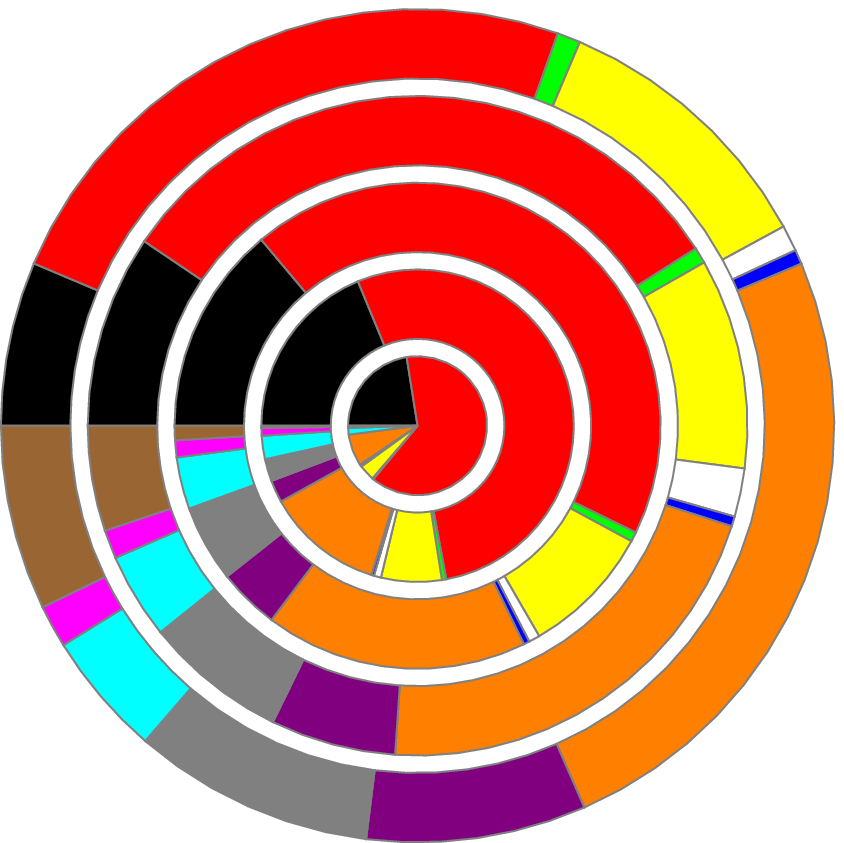} &
  \includegraphics[width=0.20\textwidth,height=!]{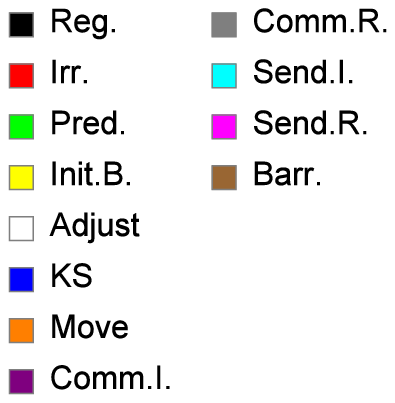} \\
  {\tiny 128k without primordial binaries} & \\
  \end{tabular}
  \caption{Pie charts showing the same test as Figure~\ref{fig:scaling} but time
    fraction of different components in Hybrid MPI parallel NBODY6++. In each
    chart different rings show different $N_{\rm node}$. From inside to outside
    rings, $N_{\rm node}$ are $1$, $2$, $4$, $8$ and $16$. The two pie charts on the left
    show the model without primordial binaries and the pie chart on the right shows the
    model with $5\%$ binaries. The models in top two pie charts include $1024k$
    particles (singles + binaries) and the model in the pie chart at the bottom
    include $128k$ single particles. An explanation of the legends is provided in
    Table~\ref{tab:def}}
  \label{fig:frac} 
\end{figure}

We show the fraction of time spent on different parts of \nbodyppgpu in Figure~\ref{fig:frac}.
In the model without binaries, MPI communication and data moving consumes about half of the total time in the case of $1024k$ particles with $N_{\rm node} = 16$ and $128k$ particles with $N_{\rm node} = 8$, 
which means the scaling reaches the MPI parallelization speed-up break-even point. 
For the $1024k$ particles with $5\%$ binaries, the KS takes about half of the calculation time when $N_{\rm node} \ge 8$. 
Thus, the KS procedures become the performance bottleneck.

\subsection{Sorting list algorithm for selecting active particles}

\begin{figure}
  \centering
  \includegraphics[width=0.5\textwidth,height=!]{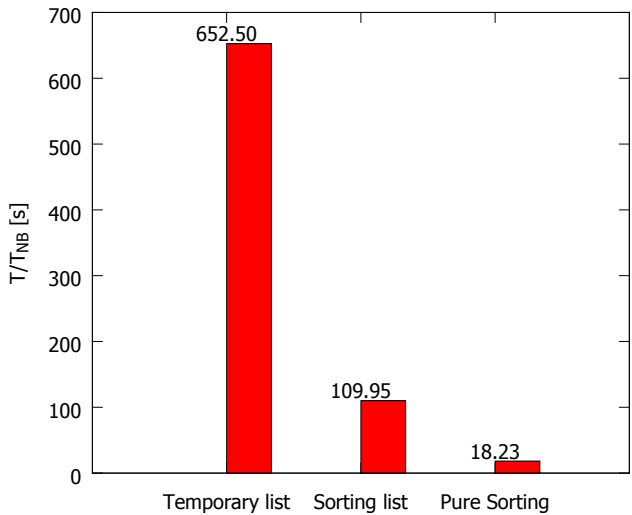}
  \caption{Comparison of performance between the sorting list algorithm and the temporary list algorithm. 
    The ``Pure Sorting'' means the time cost of sorting part in sorting list algorithm. }
  \label{fig:initblock}
\end{figure}

In Figure~\ref{fig:initblock}, we compare the performance of the sorting list algorithm and temporary list algorithm described in Section~\ref{sec:initb}.
The star cluster in our test simulation is modelled as a King sphere \citep{King1966} with $W_0 = 6$ using $1024k$ stars,  $5\%$ of primordial binaries and $8$ nodes with the same node configuration as in Figure~\ref{fig:scaling}.
To indicate that the sorting is very fast, the time of pure sorting part in this algorithm is also shown. 
We can see the sorting list algorithm is about $5$ times faster than temporary list algorithm.

The time fraction of active particle selection with the new algorithm is shown as yellow part (Init.B.) in Figure~\ref{fig:frac}. We can see Init.B. costs more for simulations with a larger number of particles. Even with this new method, it is close to irregular integration cost for the one million particles case ($\sim7\%$).

\section{Application}
\label{sec:app}
\begin{figure}
  \centering
  \includegraphics[width=0.5\textwidth,height=!]{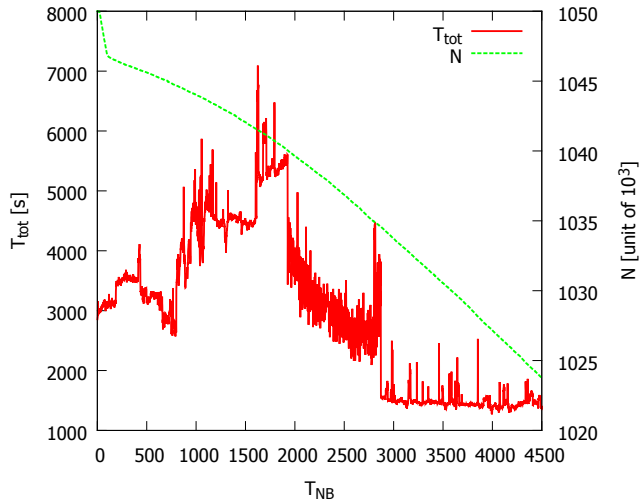}\\
  \caption{The evolution of the computing time per \nb time unit and the number of particles for the $1M$ globular cluster simulation as function of \nb time.}
  \label{fig:tevolve}
\end{figure}

The main task for \nbodyppgpu is to simulate large star clusters. 
For a typical globular cluster, the total mass is $10^5-10^6$~$M_\odot$, thus the total number of particles is of the order $10^6$. 
The typical age is about $12$~Gyr. 
In our $1M$ stars with $5\%$ primordial binary globular cluster model (the same as shown in Figure~\ref{fig:initblock}), we choose the parameters similar to NGC~$4372$ \citep{Harris1996}. 
The initial half-mass radius is $7.5$~pc and the tidal radius is $89.2$~pc with a circular orbit around a point-mass galactic potential. 
One \nb time unit corresponds to $0.622$~Myr. 
The computing time and number of particles are shown in Figure~\ref{fig:tevolve}. 

Initially, the computing time per \nb time unit was about $3000$~s and this increased when several small time step particles formed. 
Later, we carried out several adjustments, then the simulation sped up and became about $1500$~s. 
The number of particles only decreased slightly during $4500$ \nb time units, but the computing speed actually increased at a later stage. 
The reason for the early slow speed was the two unsuitable criteria for triggering or terminating the two-body KS regularization.
The first is the separation criterion $R_{\rm cl}$ and the second is the time step criterion $\Delta t_{\rm cl}$. 
If the auto-adjustment of $R_{\rm cl}$ and $\Delta t_{\rm cl}$ are used, they are determined following \cite{Aarseth2003}
\begin{equation}
  \label{eq:rtmin}
  \begin{aligned}
    R_{\rm cl} & = \frac{4 R_{\rm h}}{N (\rho_{\rm d} / \rho_{\rm h})^{1/3}}, \\
    \Delta t_{\rm cl} & \simeq 0.04 \left( \frac{\eta_{\rm I}}{0.02} \right)^{1/2} \left( \frac{R_{\rm cl}^3}{\langle m \rangle} \right)^{1/2},
  \end{aligned}
\end{equation}
where $\rho_{\rm d}/\rho_{\rm h}$ is the central density contrast, $\eta_{\rm I}$ is the standard irregular time step coefficient and $\langle m \rangle$ is average mass. 
The factor $4 R_{\rm h} / N$ is the impact parameter for a $90$ degree deflection in a two-body encounter. 
The auto-adjustment results in $R_{\rm cl} = 1.4\times 10^{-6}$ and $\Delta t_{\rm cl} = 6.8 \times 10^{-8}$ \nb units at the beginning of this simulation. 
But these values are too small and many wide binaries including some unperturbed binaries are not regularized. 
Thus we switched off auto-adjustment and used $R_{\rm cl} = 5.0\times 10^{-6}$ and $\Delta t_{\rm cl} \le 2.0 \times 10^{-7}$ before about $2800$ time units. 
We found that the $R_{\rm cl}$ and $\Delta t_{\rm cl}$ parameters were still too small, thus we enlarged $R_{\rm cl}$ to $1.0 \times 10^{-5} $ and $\Delta t_{\rm cl}$ to $5.0 \times 10^{-7}$. 
Then the computing sped up after $2800$ time units.
The small parameters from auto-adjustment is because Eq.~\ref{eq:rtmin} is originally designed for small number of particles ($N = 10^2-10^3$). 
For the million-body simulation, the central density is usually high and $\rho_{\rm d}/\rho_{\rm h}$ is large. 
The criterion from Eq.~\ref{eq:rtmin} is only suitable for the central region of the cluster but too small for the outer region.
There were several jumps in the computing time after the auto-restart with reduced time steps.
These happened when a large energy error appeared due to specific events, such as difficult triple systems or the sudden change of force caused by large mass loss or premature perturbation of the neighbor sphere (such as neutron stars with high kick velocities).
After we restore the normal time step parameters the computing time was again reduced.

There are also a few more models currently in progress and we will report in detail about the results from these simulations in a future publication.

\section{Discussion}
\label{sec:discussion}

While standard Hermite codes report a high efficiency using up to 700,000 cores on hundreds if not thousands of GPUs \citep{Berczik2013a,Berczik2013b}, we find that our performance saturates at about 86,000 GPU cores and 320 CPU cores. 
This is not surprising, because \nbody and \nbodypp are inherently more efficient than standard Hermite codes (less operations for the same physical result). 
A more detailed scaling analysis of \nbodyppgpu will be published separately (Huang et al., private communication).

As discussed in sections~\ref{sec:scale} and \ref{sec:tfrac}, the data movement and MPI communications become the bottleneck when the node number $N_{\rm node}$ is large since they have constant cost and the KS integration dominates the calculation when there are many primordial binaries.
For the data copying and communication limit, a better communication algorithm (such as non-blocking communication as suggested by \cite{Dorband2003}, which we will probably work on in the future), a higher network bandwidth between nodes and faster memory access are required. 


In the common computer architecture today, the pure calculation operations for CPU is about two orders of magnitude faster than to access data from the host memory.
For the non-shared memory parallelization like MPI, if the data communication consumption is larger than the calculation, 
the parallelization cannot improve the performance and sometimes even reduces the speed. 
Table~\ref{tab:cost} compares the calculation and communication costs for the regular force, the irregular force and the KS perturbation calculations. 
The ratio of calculation cost to communication cost, $R_{\rm c}$, for the regular force is proportional to the full particle number $N$. 
Thus, the GPU and MPI parallelization for the regular force gives a very good scaling. 
For the irregular force, $R_{\rm c}$ is proportional to the average neighbor number. 
When there are many neighbors, the MPI parallelization is good. 
For typical star cluster simulations, the neighbor number $N_{\rm b}$ is a few hundred, thus it is acceptable. 
In NBODY6++GPU, the data movement and MPI communication is significant for the irregular force (Figure~\ref{fig:frac}). 
For KS perturbation calculation, $R_{\rm c}$ is proportional to the average perturber number $N_{\rm p}$, which is usually quite small (less than $100$).
Thus, MPI parallelization for KS can be inefficient. 
The reason for the small $N_{\rm p}$ is that usually in star cluster simulations, a large fraction of the KS binaries is unperturbed with $N_{\rm p} = 0$, and perturbed KS binaries also tend to have small $N_{\rm p}$ (otherwise they would be terminated or transformed to hierarchical systems). 
Therefore, to get good performance of KS parallelization, the unperturbed and perturbed KS parts should be treated separately, since unperturbed KS only needs few operations and should avoid communication when parallelized (shared memory parallelization such as OpenMP or {\protect MPI-3}).
We are working on this and will show our KS parallelization method and benchmarks in a future publication. 
There is also another effort to parallelize KS with block time steps (Nitadori 2014, private communication).

\begin{table}
  \caption{Estimation of calculation and communication cost}
  \label{tab:cost}
  \begin{tabular}{llll}
    \hline
    Cost          & Regular force   & Irregular force             & KS perturbation \\\hline
    Calculation   & $O(N_{\rm i} N)$  & $O(N_{\rm i} \langle N_{\rm b} \rangle)$ & $O(N_{\rm i} \langle N_{\rm p} \rangle)$ \\
    Communication & $O(N_{\rm i})$        & $O(N_{\rm i})$                     & $O(N_{\rm i})$ \\\hline
  \end{tabular}\\
  *$N_{\rm i}$: Active particle number \\
  *$N$: Full particle number \\
  *$\langle N_{\rm b} \rangle$: Average neighbor number \\
  *$\langle N_{\rm p} \rangle$: Average perturber number for KS \\
\end{table}

We also find that the KS initialization and termination can be costly when there are wide binaries that frequently switch between KS and Hermite solutions.
As discussed in Section~\ref{sec:init}, during the KS initialization and termination, the force and its first three derivatives need to be renewed for center-of-mass particles or two components (cost of $O(N)$) 
and the neighbor list of every particle and perturber list of KS pairs should be updated with new particle index (cost of $O(N \langle N_{\rm b} \rangle)$). 
The regular part can be improved by using existing values instead of a direct calculation. The latter can be improved by using a reverse neighbor list for fast searching which particle has the KS pair as its neighbors. 
However, this requires large memory cost and coding effort.

When testing the code performance in computer clusters, we usually use the empty nodes where no other tasks are performed simultaneously. 
But for the applications, whether we can use scheduling whole nodes depends on the task management system in the clusters. 
Some clusters, such as ``Laohu'' at NAOC and ``Milkyway'' at the J\"ulich Computing center, only allow very few CPU cores for GPU tasks ($1-2$ CPU cores per GPU) and all other CPU cores in the same nodes are reserved for pure CPU tasks. 
\nbodyppgpu is not suitable for these kinds of clusters since it relies on heavy calculation on CPU (irregular and KS integration, data movements; see Figure~\ref{fig:frac}). 
Moreover, in the shared nodes, different tasks compete with each other for network bandwidth, CPU loading and host memory. 
This sometimes results in a serious load imbalance: 
The MPI barrier time (Table~\ref{tab:def}) covers almost half of the total computing time. 
The only solution is to use computing clusters in which GPU nodes can be fully occupied by one GPU task each time.

Both \nbody and \nbodypp have been developed over a long time. 
The codes have become more and more complicated which makes it difficult for beginners. 
Therefore, we also present documentation for the new version of NBODY6++GPU. 
The document includes a detailed description of all input parameters and output data and will be updated with more details and new implementations.
We also show several important differences between \nbodyppgpu and \nbodygpu in the Appendix.

The future improvements of the codes and hardware may lead to simulating even larger particle numbers, e.g., for nuclear star clusters using more GPU nodes appears feasible. 
The key to keep total wall clock times reasonable will be further optimization of communication and data management, especially for particles with very small time steps near a central black hole. 
Also, bandwidth and latency of communication hardware may help to gain one more order of magnitude, but not to reach the Exaflop/s regime. 
For the latter, hybrid codes seem more appropriate, which treat a large number of particles in the outskirts self-consistently, but not with full $N^2$ accuracy of the force computation (see, for recent examples, e.g., \citealp{Meiron2014,Karl2015}).

\section{Conclusions}
\label{sec:conclusion}
Direct numerical simulations of star clusters contribute significantly to the theoretical understanding of star cluster dynamics.
Due to hardware and software limits, direct \nb simulations of real globular clusters with large number of particles have been a major challenge for many years.
\cite{Sugimoto1990} pointed out that direct numerical simulations of globular star clusters could not be completed for the next decades unless there are breakthroughs in parallel computing which violate Moore's law.
After that, many efforts were made to reach this goal by using specially designed acceleration hardware (GRAPE and GPU).

In this paper, we present NBODY6++GPU.
It combines for the first time the massively parallel multi-node code (MPI parallelized) \nbodypp \citep{Spurzem1999, Hemsendorf2003} with the GPU and AVX/SSE acceleration on each node, using the libraries of \cite{Nitadori2012}. 
We discuss the performance tests (Figure~\ref{fig:mpivshybrid}, \ref{fig:scaling}, \ref{fig:regirr} and \ref{fig:frac}) and new algorithms (Figure~\ref{fig:sortchart}, \ref{fig:sortlist} and \ref{fig:initblock}) to accelerate the NBODY6++GPU. 
For the non-binary case, the overall scaling is good up to $16$ nodes ($320$ CPU cores and $32$ NIVDIA K20x GPUs including $86016$ GPU cores) with a speed up of $400$ up to $2000$ depending on the particle numbers. 
The speed up is mainly achieved by the usage of GPUs to accelerate the long-range (regular) gravitational forces, which gives about $33$ times faster force calculation (Figure~\ref{fig:mpivshybrid}). 
The AVX/SSE increase the speed of prediction of positions and velocities and neighbor (irregular) forces by a factor of $3$. 
We also worked on the consistency of the code when combining several parallel methods together to ensure the stability.
When GPU and AVX/SSE accelerate the force calculation very efficiently, other parts become bottlenecks of performance, such as time step scheduling and stellar evolution. We designed new algorithms to improve these parts.

We have demonstrated how \nbodyppgpu can simulate a realistic globular cluster with one million particles, stellar evolution and 5\% primordial binaries for several Gyr (one half-mass crossing time requiring about an hour computational time; see Figure~\ref{fig:tevolve}). A few more models are currently in progress and we will report the detailed results of these simulations in future publications.
With our final code version, which is publicly available\footnote{We use Subversion and Github to manage \nbodyppgpu. The beta version can be downloaded by commands ``svn co http://silkroad.bao.ac.cn/repos/betanb6'' or ``git clone https://github.com/lwang-astro/betanb6pp.git''}, we can claim to have finally reached the goal of Sugimoto's ``dream'' of 1990. 
A million-body cluster can be simulated for about $20$ crossing times in one day on $320$ cores with $32$ GPUs.
In the future, with the faster bandwidth and latency of hardware as well as optimizations of communication and data management, even larger system like the nuclear star clusters may be simulated by direct \nb codes.

The previous paragraphs show that our contribution to this would be impossible without the achievements of our predecessors and collaborators; in particular the current dominance of GPU hardware has been assisted by the development of GRAPE software over the last few decades which finally could be ported to GPU without fundamental problems.

\section*{Acknowledgments}
This work has been partly funded by National Natural Science Foundation of China, No. 11073025 (RS).
We acknowledge support through the Silk Road Project at National Astronomical Observatories of China (NAOC, http://silkroad.bao.ac.cn).

R.S. and P.B are grateful for support by the Chinese Academy of Sciences Visiting Professorship for Senior International Scientists, Grant Number 2009S1$-$5, and through the ``Qianren'' special foreign experts program of China, both at NAOC.

Most of the numerical simulations have been done on the ``Hydra'' GPU cluster of the Max-Planck Supercomputing Centre (RZG) Germany.

R.S. and P.B. and L.W. are grateful for kind hospitality and support during several visits at the Max-Planck-Institute for Astrophysics.
Other resources used for numerical simulations in the preparation of this paper are:
``Laohu'' GPU cluster at the Center of Information and Computing at NAOC, ``Kepler'' GPU cluster at ARI/ZAH, University of Heidelberg, Germany (funded by Volkswagen Foundation) and the ``MilkyWay'' cluster of SFB 881 “The Milky Way System” at the University of Heidelberg, Germany, hosted and co-funded by the J\"ulich Supercomputing Center (JSC).

S.A. and K.N. are grateful for support during their visits at Kavli Institute for Astronomy and Astrophysics, Peking University and NAOC.

P.B. acknowledges the special support by the NASU under the Main Astronomical Observatory GRID/GPU computing cluster project. 

M.B.N.K. was supported by the Peter and Patricia Gruber Foundation through the PPGF fellowship, by the Peking University One Hundred Talent Fund (985), and by the National Natural Science Foundation of China (grants 11010237, 11050110414, 11173004). 
This publication was made possible through the support of a grant from the John Templeton Foundation and NAOC.
The opinions expressed in this publication are those of the author(s) do not necessarily reflect the views of the John Templeton Foundation or NAOC. 
The funds from John Templeton Foundation were awarded in a grant to The University of Chicago which also managed the program in conjunction with NAOC.

TN acknowledges support by the DFG cluster of excellence ‘Origin and Structure of the Universe’.

We thank the anonymous referee for many useful comments that helped to improve the paper.



\onecolumn
\appendix
\section{Differences between \nbodypp and \nbody}
There are several differences with \nbody. 
Table~\ref{tab:diff} lists some of the most important. 
The manual in the \nbodyppgpu code directory gives more details.
\newline

\begin{table*}
  \label{tab:diff}
  \caption{Differences between NBODY6++ and NBODY6}
  \begin{tabular}{p{0.8in}|p{0.8in}|p{2.1in}|p{2.1in}}
    \hline
    & Subroutine &\nbodypp & \nbody\\
    \hline
    Installation  &     
                  & Use configure script (see GPU Autoconf software$^{1}$) 
                  & Use Makefile \\\hline
    Parallelization &
                    & Can enable/disable features among MPI, GPU, OpenMP and AVX/SSE,  except AVX/SSE requires OpenMP enabled
                    & Use GPU, OpenMP and AVX/SSE together or OpenMP with AVX/SSE (only OpenMP or GPU with OpenMP are not supported) \\\hline
    Data files    &     
                  & Rename most of output data files, change contents of some files and describe all data in a manual 
                  & Describe in a document \\\hline
    Basic data initialization & 
                              & Support different reading data format (see the manual for option KZ(22)) 
                              & Support \nb and astronomical unit data format \\\hline
    Primordial binary initialization & binpop.[f/F] 
                                     & Support period distribution \citep{Kroupa1995} 
                                     & Support modified period distribution with maximum semi-major axis $1000$ AU and mimimum period $1$ day \\\hline
    Neighbor criterion & regcor\_gpu.f gpucor.f 
                       & Adjust neighbor number to input parameter \texttt{NNBOPT} 
                       & Adjust neighbor number based on density contrast \\\hline
    Stellar evolution  & kick.[f/F]
                       & Use Maxwellian distribution of neutron star kick velocity with velocity dispersion $265$ km/s (one dimension; \citealp{Hobbs2005}) 
                       & Use Maxwellian distribution of kick velocity with velocity dispersion $2\times$ \texttt{VSTAR} (velocity scaling factor) and maximum kick velocity $10\times$ \texttt{VSTAR} \\
                       & hrdiag.f kick.[f/F]
                       & Use mass fall back for black hole kick (increase the remnant mass and reduce kick velocity; \citealp{Belczynski2002}) 
                       & Use black hole mass based on \cite{Eldridge2004} (may add \citealp{Belczynski2002} kick method in the future)\\
                       & brake4.f & -- & Support gravitational radiation analytical orbit shrinkage\\
                       & intgrt.[f/F] intgrt\_omp.f 
                       & Apply mass loss only during regular step for thread-safety
                       & Apply mass loss with minimum time step $100$ years\\
                       & mdot.[f/F] 
                       & Only apply force correction when large mass loss happens (less accuracy but much faster)
                       & Calculate new force and its derivates for large mass loss\\\hline
    Galactic tidal force & xtrnlf.f fbulge.f
                         & Support point-mass + disk + halo + Plummer model 
                         & Support point-mass + disk + halo + bulge + Plummer model\\\hline
  \end{tabular}\\
  $^{1}$ \protect{http://www.gnu.org/software/autoconf/}
\end{table*}

\end{document}